# Simple ansatz for the lattice fermion determinant

Sergei V. Zenkin[1]

Laboratoire de Physique Théorique et Hautes Énergies[2]
Université de Paris XI, 91405 Orsay Cedex, France

### Abstract

An ansatz for the fermion vacuum functional on a lattice is proposed. It is proved to reproduce correct continuum limit for convergent diagrams of any finite order in smooth external fields, as well as consistent chiral anomalies, and ensures gauge invariance of the absolute value of the functional at any lattice spacing. The ansatz corresponds to a certain non-local fermion action having global chiral invariance. Problems caused by non-smooth gauge fields are discussed.

---

[1]Permanent address: Institute for Nuclear Research of the Russian Academy of Sciences, 117312 Moscow, Russia. E-mail address: zenkin@al20.inr.troitsk.ru
[2]Laboratoire associé au Centre National de la Recherche Scientifique

# 1 Introduction

The more symmetries of the target theory a regularization retains, the better it is. Chiral symmetry is a stumbling-block for any regularization, and the lattice regularization, being the basis for the most powerful nonperturbative methods, is not an exception [1]. There are deep reasons for this, which are intimately related to the existence of chiral anomalies [2]. This causes the well-known problems with definition of chiral gauge theories on a lattice (for a review of the recent approaches and the references see [3]), where the main aim is to formulate the anomaly-free theories in such a way, that both real and imaginary parts of the fermion vacuum functional $\exp\Gamma$, $\Gamma = \Re\Gamma + i\Im\Gamma$, would be gauge invariant without introducing additional counterterms and tuning there parameters.

There have been proposed several such formulations [4, 5, 6] (for a more complete list of references, see [3]). In this paper we propose an ansatz for $\Gamma$ which is free of some of the drawbacks of these formulations and has most of their advantages. For instance, it does not involve an infinite set of degrees of freedom (unlike [4]), ensures gauge invariance of $\Re\Gamma$ at any lattice spacing, and in the continuum limit for smooth external fields reproduces consistent anomalies and yields the gauge invariant non-anomalous part of $\Im\Gamma$ (like the overlap [5] and 'hybrid' [6] formulations). Besides, it is generated by a certain fermion action retaining global chiral invariance (unlike the overlap and 'hybrid' formulations).

The ansatz introduces, in fact, a minimum violation of the gauge invariance. Simple analytic estimate of the gauge variation of $\Im\Gamma$, however, demonstrates that even such a mild violation of the gauge invariance may become strong when the gauge fields are not smooth. In the full theory (where the gauge fields are dynamical) suppression of such a violation may require smoothing the gauge fields (see [3, 6] and references therein), which appears to be necessary in all other formulations.

In Section 2 we introduce the ansatz. Its basic properties are explained in Section 3. In Section 4 we examine it for smooth, and in Section 5 for non-smooth external fields. Section 6 is a short conclusion.

Our conventions are the following. We consider Euclidean hypercubic $D$-dimensional lattice with spacing $a$ and volume $V = (aN)^D$, where $D = 2$ or $4$ and $N/2$ is even. Topologically, the lattice is a torus $T^D$; its sites numbered by $D$-dimensional vectors $n$; $\hat{\mu}$ are unit vectors along positive directions. Dirac $\gamma$-matrices are hermitian, $P_L = (1 + \gamma_5)/2$. Fermion (boson) variables obey antiperiodic (periodic) boundary conditions for all directions.

## 2 Ansatz

We consider a lattice regularization of the normalized fermion vacuum functional

$$\exp \Gamma[A] = \frac{1}{\mathcal{N}} \int \prod_n d\psi_n d\overline{\psi}_n \exp \sum_{m,n} \overline{\psi}_m D_{mn}(A) \psi_n$$
$$= \exp \operatorname{Tr} \ln[D(A) D^{-1}(0)], \tag{1}$$

where $D(A)$ is a lattice transcription of the chiral Dirac operator $\gamma_\mu(\partial_\mu + ig A_\mu P_L)$.

The ansatz is based on the observation [7] that the gauge invariance of $\Re\Gamma[A]$ without species doubling can be achieved if in the naive formulation the domain of integration over fermion loop momenta in all diagrams is narrowed down to $\mathcal{D} = (-\pi/(2a), \pi/(2a))^D$, i.e. to the $1/2^D$-th part of the fermion Brillouin zone $\mathcal{B} = (-\pi/a, \pi/a)^D$. In this case the fermion modes which lead to the species doubling and render any theory vector-like [8, 1] are no longer dangerous, for they almost decouple from the smooth fields though are still important for restoring the gauge invariance of $\Re\Gamma[A]$.

Our ansatz realizes such a procedure, and reads as follows:

$$\Gamma[A] = \operatorname{Tr}\{\Theta \ln[\nabla(U) \nabla^{-1}(1)]\}, \tag{2}$$

where $\nabla(U)$ is the naive lattice transcription of the Dirac operator,

$$\nabla_{mn}(U) = -\sum_\mu \gamma_\mu \frac{1}{2a} (U_{m\,m+\hat{\mu}} \delta_{m+\hat{\mu}\,n} - U_{m\,m-\hat{\mu}} \delta_{m-\hat{\mu}\,n}),$$
$$U_{m\,m\pm\hat{\mu}} = \exp[\pm ig a A_\mu(m \pm \hat{\mu}/2) P_L], \tag{3}$$

and $\Theta$ is the projection operator that cuts out the proper $1/2^D$-th part of the Brillouin zone in the fermion loop integrals[3]:

$$\Theta_{mn} = \frac{1}{V} \sum_{p \in \mathcal{B}} \exp[ip(m-n)a] \prod_\mu \Theta(p_\mu),$$
$$\Theta(p_\mu) = \begin{cases} 1 & \text{if } p_\mu \in (-\pi/(2a), \pi/(2a)) \pmod{2\pi/a}, \\ 0 & \text{otherwise.} \end{cases} \tag{4}$$

---

[3]The integrals over fermion and boson momenta, $\int_\mathcal{B} d^D p/(2\pi)^D$ and $\int_\mathcal{B} d^D q/(2\pi)^D$, are defined as limits at $N \to \infty$ of the finite sums $V^{-1} \sum_{p \in \mathcal{B}}$ and $V^{-1} \sum_{q \in \mathcal{B}}$ respectively.

The explicit form of $\Theta$ in the position space is

$$\Theta_{mn} = \frac{1}{N^D} \prod_\mu \frac{\sin[\pi(m_\mu - n_\nu)/2]}{\sin[\pi(m_\mu - n_\nu)/N]} \tag{5}$$

for $m \neq n$, and $\Theta_{m\,n=m} = 1/2^D$.

Formally, the ansatz corresponds to the non-local lattice Dirac operator

$$D(A) = \exp\{\Theta \ln[\nabla(U)\nabla^{-1}(1)] + \Theta \ln \nabla(1)\}, \tag{6}$$

that retains global chiral invariance of the fermion action in (1). Expression (2) has also the following constructive representation:

$$\text{Tr}\{\Theta \ln[\nabla(U)\nabla^{-1}(1)]\} = \int_0^1 dt\, \text{Tr}\{\nabla(U-1)\Theta\,[\nabla(1) + t\nabla(U-1)]^{-1}\}, \tag{7}$$

where $t \in [0,1]$ is a real parameter. The right-hand side of (7) has a simple meaning: its integrand is the sum of all fermion loop diagrams with the operator $\Theta$ inserted into one of the vertices, while the integration over $t$ restores the proper weights of these diagrams. Note, that in the r.h.s. of (7) only local matrices are inverted, so the main difficulty is brought by the integration over $t$.

Let us now demonstrate the basic properties of our ansatz.

## 3 Basic properties

In the formal expansion of $\Gamma[A]$ in $gA$ the term of the order of $n$ in $gA$, $\Gamma_n[A]$, is represented by fermion loop diagrams with $n$ external legs. Note, that fermion propagator, as well as all vertices, in our formulation are exactly the same as in the naive formulation: they are sines or cosines of the corresponding momenta, and the only difference is the insertion of the operator $\Theta$ into the fermion loops. In particular, the Ward identities for the vertices, which follow from gauge invariance of the naive formulation [8], are unaffected by $\Theta$.

Since our formulation maintains global chiral invariance, each diagram is decomposed into a sum of two terms differing from each other only in the presence or absence of $\gamma_5$ under the trace sign of the corresponding expressions. The term without $\gamma_5$ is real and contributes to $\Re\Gamma$. The term with $\gamma_5$ is imaginary and contributes to $\Im\Gamma$. Note that the denominator of the integrand of any diagram is $\pi$-periodic function of the loop momentum.

Consider the real part of a diagram. Since the trace of the product of $\gamma$-matrices in this case is the sum of products of Kronecker's $\delta$-symbols, all the components of the loop momentum appear in the numerator of the integrand in $\pi$-periodic combinations like $\cos(p_\mu a + \cdots)\sin(p_\mu a + \cdots)$, etc. Therefore the integrand is a $\pi$-periodic function of the loop momentum. From here it follows the well-known result [8] that the Brillouin zone in the naive formulation is naturally broken into $2^D$ equal domains, each giving the same — gauge invariant — contribution to $\Re\Gamma[A]$. Our ansatz picks out only the contribution of one of these domains (namely, of $\mathcal{D}$) and therefore yields a gauge invariant result for $\Re\Gamma[A]$, which, taking into account the relationship between the naive and staggered fermions [9], is

$$\Re\Gamma[A] = \frac{1}{2^D}\Gamma_{naive}[A] = \frac{1}{2^{D/2}}\Re\Gamma_{staggered}[A]. \qquad (8)$$

The situation is quite different for the imaginary part of the diagram. The trace now includes $\gamma_5$ and results in the sum of terms each of which involves the antisymmetric tensor $\epsilon$, so all the components of the loop momentum appear in the numerator of the integrand in $\pi$-antiperiodic combinations like $\epsilon_{\mu\nu\ldots}\left[\cos(p_\mu a + \cdots)\sin(p_\nu a + \cdots)\cdots\right] \times (\pi\text{-periodic function})$, etc. Therefore the integrand now is $\pi$-antiperiodic function of the loop momentum. In this case in the naive formulation the above subdomains of the Brillouin zone give the contributions to the $\Im\Gamma[A]$ which are the same only modulo sign, so that in sum one has $\Im\Gamma_{naive}[A] = 0$.

Let us show that this leads to gauge noninvariance of $\Im\Gamma[A]$ and that this noninvariance has a simple origin very similar to what one has in the continuum theory. Consider all the diagrams of $n$th order in the external fields, and make an infinitesimal gauge transformation of $A$, $A^\omega$. Making use of Ward's identities, the expression for the gauge variation of $\Gamma_n[A]$, $\delta\Gamma_n[\omega, A] = \Gamma_n[A^\omega] - \Gamma_n[A]$, can be presented in terms of differences of two momentum integrals over the domain $\mathcal{D}$ with the integrands differing from each other by shifts of the loop momentum[4]. Due to periodicity of the real parts of the integrands in the domain $\mathcal{D}$ one can make appropriate shifts of the integration variable that results in $\delta\Re\Gamma_n[\omega, A] = 0$ for any $n$. For the imaginary parts, $\mathcal{D}$ is no longer the period of the corresponding integrands and such shifts result in the appearance of a kind of surface terms. It is well known however (see, for example, [10]) that such terms vanish when the regulator is removed, provided the shifts are finite and the corresponding integrals converge or at most diverge logarithmically. This is exactly what happens in the case of the diagrams with $n > D$, while non-vanishing diagrams with $n \leq D$ give rise to anomalies.

---

[4]Such shifts are linear combinations of the external momenta.

Indeed, for infinitesimal gauge transformations the quantity $\delta\Im\Gamma_n[\omega, A]$ has the form

$$\delta\Im\Gamma_n[\omega, A] = \int_{\mathcal{B}} \frac{d^D q_1}{(2\pi)^D} \cdots \frac{d^D q_{n-1}}{(2\pi)^D} \, \text{tr}[\omega(-q_1 - \cdots - q_{n-1}) \\ \times A_{\mu_1}(q_1) \cdots A_{\mu_{n-1}}(q_{n-1})] \, \delta\Im\Gamma_{\mu_1\cdots\mu_{n-1}}(q), \tag{9}$$

and simple estimates show that

$$\delta\Im\Gamma_{\mu_1\cdots\mu_{n-1}}(q) = O(a^{n-D}(q_1 + \cdots + q_{n-1})). \tag{10}$$

## 4 Smooth external fields

We still should convince ourselves that the continuum limit of $\Gamma_n[A]$ defined by our ansatz is not in contradiction with what one has in the continuum perturbation theory. That this is indeed so follows from the power counting arguments, which for $n > D$ can be substantiated rigorously. For this it is sufficient to note that for any finite $n$ and fixed external momenta $q_1, \cdots q_n$ there exists such a positive number $a_n$, that for any $a < a_n$ all conditions of Reisz's power counting theorem for one loop [11] are satisfied. This guarantees the correct continuum limit for $\Gamma_{n>D}[A]$ for such $A$ that $\lim_{a\to 0} A(q) = 0$ for any finite $qa$, i.e. for smooth external fields. In particular, for such $A$ the continuum limit of $\Im\Gamma_{n>D}[A]$ is gauge invariant, the result that also follows from (10).

As concerns the contributions with $n \leq D$, we shall restrict ourselves to the direct calculations of $\Gamma_2[A]$ for $D = 2$ and of $\delta\Im\Gamma_3[\omega, A]$ for $D = 4$ at smooth $A$.

In the two-dimensional theory $\Gamma_2[A]$ has the form

$$\Gamma_2[A] = -\frac{1}{2} \lim_{N\to\infty} \frac{1}{(aN)^2} \sum_{q\in\mathcal{B}} \text{tr}[A_\mu(-q) A_\nu(q)] \\ \times [\Re\Pi_{\mu\nu}(qa, N) + i\Im\Pi_{\mu\nu}(qa, N)], \tag{11}$$

where the trace is taken over the gauge group indices and $\Pi(qa, N)$ is determined by the fermion loops with two external legs. By direct computation of $\Pi(qa, N)$ for increasing $N$ and $qa = o(N)$ we find that

$$\Re\Pi_{\mu\nu}(qa, N) \to \frac{g^2}{2\pi}(\delta_{\mu\nu} - \frac{q_\mu q_\nu}{q^2}),$$
$$\Im\Pi_{\mu\nu}(qa, N) \to \frac{g^2}{4\pi}\frac{1}{q^2}(\epsilon_{\mu\alpha}q_\alpha q_\nu + q_\mu \epsilon_{\nu\alpha}q_\alpha), \tag{12}$$

and $\Re\Pi_{\mu\,\nu=\mu}(0) = g^2/(4\pi)$, $\Re\Pi_{\mu\,\nu\neq\mu}(0) = 0$, and $\Im\Pi_{\mu\nu}(0) = 0$. For the abelian theory this is the well-known exact result for $\Gamma[A]$ [12]. From the above expressions directly follows the result for $\delta\Im\Gamma_2[\omega, A]$:

$$\delta\Im\Gamma_2[\omega, A] = i\frac{g^2}{4\pi} \int \frac{d^2q}{(2\pi)^2} \text{tr}[\omega(-q)A_\mu(q)]\epsilon_{\mu\alpha}q_\alpha. \tag{13}$$

In a similar way, in the four-dimensional theory we find for smooth $A$

$$\delta\Im\Gamma_3[\omega, A] = \frac{g^3}{24\pi^2} \int \frac{d^4q_1}{(2\pi)^4} \frac{d^4q_2}{(2\pi)^4} \text{tr}[\omega(-q_1-q_2)A_{\mu_1}(q_1)A_{\mu_2}(q_2)]\epsilon_{\mu_1\mu_2\alpha\beta}q_{1\alpha}q_{2\beta}. \tag{14}$$

Thus, for smooth external fields our ansatz reproduces the consistent anomaly [13] for the two-dimensional theory, and at least its triangle part for the four-dimensional theory.

## 5  Non-smooth external fields

For smooth external fields our ansatz looks almost perfect. The problems arise when one removes this limitation, or, equivalently, when the external momenta $q$ in the diagrams are no longer kept finite at $a \to 0$. It is clear that the properties of $\Gamma[A]$ at any $A$ become important in the full theory, where functional integration is performed over the gauge degrees of freedom as well. There are two potential problems.

The first one is that in our formulation the fermion modes of opposite chirality are still present, and their interactions are suppressed only for the smooth fields[5]. In particular, because of this the Reisz theorem [11] is inapplicable in our case for arbitrary $q$, and, therefore, beyond one loop. To get some idea of what may happen at high external momenta, we return to direct calculations in the two-dimensional model.

Consider the difference

$$\Delta_{\mu\nu}(qa, N) = \frac{2\pi}{g^2}\Re\Pi_{\mu\nu}(qa, N) - (\delta_{\mu\nu} - \frac{q_\mu q_\nu}{q^2}) \tag{15}$$

[see eq. (12)]. In a perfect case $\Delta_{\mu\nu}(qa, N)$ should be zero at any $q \in \mathcal{B}$. Normally, for a gauge invariant formulation we would expect that it is almost

---

[5]This is the point at which our ansatz differs essentially from the Zaragoza proposal [14], where the coupling of these modes to the gauge fields is suppressed at any momenta. However, the price of such a suppression is the breaking of gauge invariance for the real part of $\Gamma[A]$.

zero within a central region, whose size depends on the quality of the formulation, with some smooth deviations from zero closer to the boundary of the Brillouin zone[6]. And this is indeed the case for $\Delta_{\mu\,\nu\neq\mu}$. The situation is worse for $\Delta_{\mu\,\nu=\mu}$, see Fig. 1 which shows $\Delta_{00}$ for $N = 32$. At $N \to \infty$ the quantity

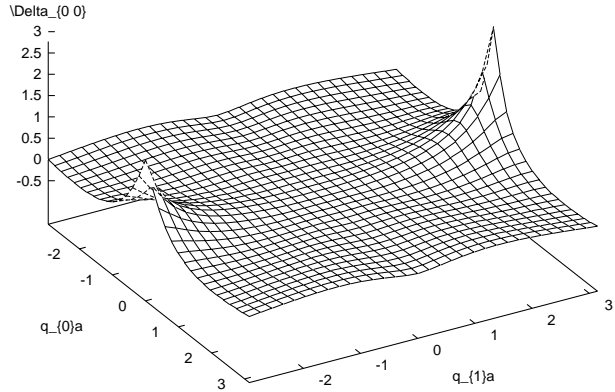

Figure 1: $\Delta_{00}(qa, N)$ of eq. (15) at $N = 32$.

$\Delta_{00}(q_0 a = 0, q_1 a = \pi, N)$ diverges logarithmically, and we expect in this limit that

$$\Delta_{00}(qa, N) \to c\,\delta(q_0 a)\,\delta(q_1 a - \pi) + \text{smooth terms}, \qquad (16)$$

where $c$ is a positive constant. The situation is similar for $\Delta_{11}$ (with $q_0$ and $q_1$ interchanged). Simple perturbative considerations, however, show that such features do not affect the results beyond one loop. Indeed, $\Pi(q)$ always appears in higher-order diagrams between two gauge field propagators, each of which is $O(q^{-2})$, so the above singularity gives no contribution to such diagrams. Another — indirect — argument in favour of such a conclusion, which holds in four dimensions as well, is that exactly the same thing happens with staggered fermions [up to a factor $2^{D/2}$ in front of $\Gamma[A]$, see eq. (8)]. The latter are known to be an adequate lattice transcription of Kähler's fermions [15], which in their turn are $2^{D/2}$-plets of Dirac fermions. Thus, as concerns the real part of $\Gamma[A]$, no pathologies should appear. Even more so, in the central region of the Brillouin zone our ansatz leads to much better convergence of $\Re\Pi(qa,N)$ to the continuum expression than do the Wilson fermions. For instance, for $\Delta_{00}(qa,N)$ at $N=32$ (Fig. 1) the central square where the deviation from zero does not exceed $\pm 0.1$ contains 225 points, while for the Wilson fermions it contains only 25 points.

---

[6]This deviation can be made smoother if in the r.h.s. of eq. (15) one takes the quantity $\hat{q}_\mu = \frac{1}{2}\sin(\frac{1}{2}q_\mu)$ instead of $q_\mu$. I am grateful to Ph. Boucaud for drawing my attention to this point. We prefer, however, to keep the original $q$ in (15) to be able to estimate the deviation from the perfect case.

The imaginary part of $\Gamma[A]$ is also sensitive to the presence of these higher modes. Namely, they make all the contributions to $\Im\Gamma[A]$ vanishing at momenta $q$ going to the boundary of the Brillouin zone. In particular, they make the anomalies in our formulation a "low-energy phenomenon". To illustrate this we compute the following ratio:

$$R_\nu(qa,N) = \frac{4\pi}{g^2} \frac{\sum_\mu \frac{1}{2}\sin(\frac{1}{2}q_\mu)\Im\Pi_{\mu\nu}(qa,N)}{\epsilon_{\nu\alpha}q_\alpha}. \qquad (17)$$

In the above ideal case, $R_\nu(qa,N)$ would be unity for any $q \in \mathcal{B}$ (at those points where $R$ is indefinite we define it by continuity). The ansatz yields the picture presented in Fig. 2, where $R_0$ is shown for $N=32$.

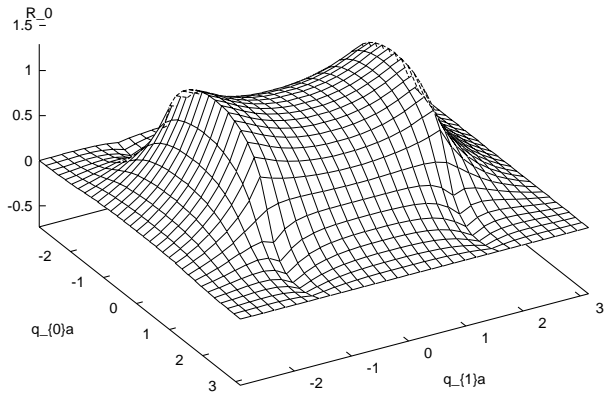

Figure 2: $R_0(qa,N)$ of eq. (17) at $N=32$.

The second problem of our formulation is common to all formulations reproducing the anomalies: even in anomaly-free models the imaginary part of $\Gamma[A]$ is not gauge invariant at finite lattice spacing. As it is seen from eq. (10), at $q = O(1/a)$ the gauge variation of the contribution of the $D+1$th order to the effective action is no longer suppressed by a power of $a$. In our formulation the problem is soften by the vanishing of all the contributions to $\Im\Gamma[A]$, and therefore of the variation (10), near the boundary of the Brillouin zone. However, due to the properties of the trigonometric functions one can expect that they drop sufficiently fast at $q > f/a$, where $f$ is some fraction of $\pi$ (see, for instance, Fig. 2). So the region $q \leq O(f/a)$ may be still dangerous.

# 6 Conclusion

The question of the violation of gauge invariance by non-smooth gauge fields deserves further investigation, since it determines whether in the full theory one should smooth the gauge fields on the scale $a$. The smoothing would mean the introduction in the theory an additional scale $b \gg a$ which limits the domain of changing of the external momenta $q$ to $|q| \leq O(1/b)$. Then, as it follows from (10), in the absence of anomalies the gauge non-invariance of $\Im\Gamma[A]$ is controlled by the ratio $r = a/b \ll 1$. The procedure of the smoothing may consist in imposing constraints on the gauge variable measure or in defining the gauge variables on the sublattice with the spacing $b$ with subsequent their interpolation to the original lattice (see [3, 6] and references therein). So, in this case the part of the problems of the chiral gauge theories moves to the gauge sector.

# 7 Acknowledgements

I am grateful to Ph. Boucaud, S. Randjbar-Daemi, A. A. Slavnov, and R. L. Stuller for interesting discussions. It is a pleasure to thank Ph. Boucaud and LPTHE in Orsay, S. Randjbar-Daemi and ICTP, and E. Seiler and MPI in Munich, where parts of this work have been done, for their hospitality. This work was partly supported by the Russian Basic Research Fund under the grant 95-02-03868a and by the Programme PECO du Ministère Français de l'Enseignement Supérieur et de la Recherche.